\documentclass[11pt]{article}
\usepackage{moriond}

\usepackage{hhline}
\def\Journal#1#2#3#4{{#1} {\bf #2}, #3 (#4)}

\def\be{\begin{equation}}
\def\ee{\end{equation}}
\def\bea{\begin{eqnarray}}
\def\eea{\end{eqnarray}}

\newcommand{\Photo}{}

\begin{document}
ULB-TH/14-08
\vspace*{4cm}
\title{NEUTRAL AND MILLICHARGED DARK MATTER DECAY INTO GAMMA-RAY LINES}

\author{ T. SCARN\`A }

\address{Service de Physique Th\'eorique, Brussels University (ULB), 2 Bld du Triomphe,\\
Brussels 1050, Belgium}

\maketitle\abstracts{
A gamma-ray line observation would be a strong hint towards the detection of dark matter. The possibility that a decay of the dark matter particle is responsible for the emission of a line is investigated, for both a neutral and a millicharged  dark matter. We focus here on a comparison of these two scenarios, based on an effective field theory description and relying on  cosmic ray continuum constraints.}

\section{Introduction}
Even though its presence is well established, the nature of dark matter (DM)  as a particle remains mysterious. The detection of monochromatic photons in the GeV energy range, producing a specific feature in the spectrum referred to as gamma-ray line, is a  so-called "smoking-gun" signature of DM, as no astrophysical process is expected to mimic such signal.  Current instruments like the Fermi-LAT \cite{fermi} and HESS \cite{hess}, but also forthcoming  Cherenkov Telescopes \cite{CTA} are or will be probing this possibility.

In contrast with the case of an annihilation, there is no link between the DM relic abundance and its partial decay length into photons, so in principle the intensity of a line might be stronger for a decay than for an annihilation. DM has to be long-lived, but it could be metastable rather than stable. If the DM stability is due to an accidental low energy symmetry, this is a natural outcome.  The ultraviolet (UV) theory might indeed break this symmetry and cause a decay suppressed by powers of the UV scale, as expected for the proton in some extensions of the SM. At low energy, this can be parametrized in full generality by writing down the most general effective theory respecting the low energy symmetries of the model.

A necessary condition for the DM to be able to produce $\gamma$-ray lines is the existence of a coupling between the DM and the photon, which can either be direct or loop-induced. The latter possibility is the most studied one, because in the majority of models, DM is assumed to be exactly neutral. In this case, the coupling to the photon has to be mediated by a loop of a charged particle.  The other option, a direct coupling to the photon, is much less explored, but it is in principle viable if the DM is millicharged.  Such tiny charge could be put by hand, but more interestingly, there are scenarios in which DM could acquire such a millicharge dynamically \cite{stueck,kinmix}.  The most general effective theory corresponding to the decay of neutral DM \cite{neutralDDM} and millicharged DM \cite{milliDDM} has been worked out, and constraints on the operators of the effective field theory have been derived.  In Sec. \ref{EFT} , we will briefly remind the structures of the operators in both cases, and in Sec. \ref{constraints} we will focus on a  comparison of the resulting constraints. 

\section{Effective Field Theory}
\label{EFT}
In order for an effective approach to provide an accurate description of the low-energy physics, the scale at which it is used must be clearly separated from the scale at which a UV completion is expected. In the case of the DM decay into  monochromatic photons, there are bounds on the lifetime of the order of $\tau_{DM \rightarrow \gamma X} \simeq 10^{28}$s \cite{fermi}. Considering for example a dimension-six operator, the lifetime typically goes as $\tau_{DM \rightarrow \gamma X} \simeq \Lambda^4_{UV}/M^5_{DM}$,
meaning that working at the electroweak scale with $M_{DM} \simeq 100$ GeV, the required UV scale compatible with the observational bound is $\Lambda_{UV}\simeq 10^{16}$ GeV, so the use of the effective approach  is fully justified here. In the case of millicharged DM, $\tau_{DM \rightarrow \gamma X}$ is boosted by the DM millicharge with  a factor of $1/ Q^2_{DM}$ with respect to the neutral case, bringing $\Lambda_{UV}$  down to $10^8$ GeV when considering values of $Q_{DM}$ close to the experimental bounds,  and  again the use of the effective theory is enabled. 

Writing down the most general effective field theory requires to make the complete list of all operators respecting the low-energy symmetries, namely here the SM gauge group. In order to get a complete but minimal set of operators, equations of motions have to be taken into account, because they allow to relate operators among them. In the particular case of DM decay into a $\gamma$-ray line, the operators must lead to a two-body decay including at least a photon,  the number of involved  particles is therefore  limited and so is the number of operators. Allowing  another beyond the SM particle  on top of the DM, the total number of operators up to dimension six  sums up to thirty in the case of neutral DM, and to thirteen in the case of millicharged DM. 

\subsection{Neutral DM}
\begin{table}[h]
\caption[]{Operators structures in the case of neutral DM decay into $\gamma$-ray line. Operators with a scalar inside parenthesis are either a  dimension-five operator when  the scalar into parenthesis is omitted or a  dimension-six operator when the scalar is included.}
\label{tab:neutral}
\vspace{0.4cm}
\begin{center}
\begin{tabular}{|c c c|}
\hline
Fermion DM & Scalar DM & Vector DM \\
\hline
 $\bar{\psi} \sigma_{\mu \nu} \psi_{DM}F^{\mu \nu} (\phi)$ &$\phi_{DM}F_{\mu \nu} F^{\mu \nu} (\phi)$& $F^{DM}_{\mu \nu} F^{Y \nu \rho} F''^{\rho}_{\mu}$ \\
 $D_\mu \bar{\psi} \gamma_\nu \psi_{DM} F^{\mu \nu}$ &$\phi_{DM}F_{\mu \nu} F^{\mu \nu} (\phi)$ & $F^{DM}_{\mu \nu} F^{\mu \nu} \phi (\phi')$ \\
 $\bar{\psi} \gamma_\mu D_\nu \psi_{DM} F^{\mu \nu}$ & $D_\mu \phi_{DM} D_\nu \phi F^{\mu \nu}$ &  $D_\mu^{DM}\phi D_\nu^{DM} \phi' F^{\mu \nu}$ \\
\hline
\end{tabular}
\end{center}
\end{table}
The thirty operators of the complete list can be sorted out in structures listed in Tab.\ref{tab:neutral}.  A necessary condition for the operators related to neutral DM to include a photon is to contain a field strength tensor as the photon cannot be included in a covariant derivative acting on a neutral field. So each operator can be declined in a version including  the hypercharge field strength and another one including the field strength of the $SU(2)_L$ gauge group. Furthermore, each operator will have a different phenomenology depending on the SM quantum numbers of the fields it contains. For example, a different amount of cosmic rays  (CR) will be associated to different couplings to the Z and W bosons. This aspect is studied in more details in Sec. \ref{constraints}.

\subsection{Millicharged DM}
\begin{table}[h]
\caption[]{Operators structures in the case of millicharged DM decay into $\gamma$-ray line. Same comment about the scalar than in Table \ref{tab:neutral}.}
\label{tab:milli}
\vspace{0.4cm}
\begin{center}
\begin{tabular}{|c c c|}
\hline
Fermion DM & Scalar DM & Vector DM \\
\hline 
 $D_\mu D_\nu\bar{\psi} \sigma_{\mu \nu} \psi_{DM} (\phi)$  &$\phi_{DM}F^A_{\mu \nu} F^{A \mu \nu} (\phi)$& $\phi F^A_{\mu \nu} F^{A \mu \nu} (\phi') $ \\
 $\bar{\psi} \sigma_{\mu \nu} D_\mu D_\nu \psi_{DM} (\phi)$ &$F^{A \mu \nu} D_\mu \phi_{DM} D_\nu \phi'$ & $F^{A \mu \nu} D_\mu \phi  D_\nu \phi' $ \\
 $D_\mu \bar{\psi} \sigma_{\mu \nu}  D_\nu \psi_{DM} (\phi)$ & &  $F^{A\mu \nu} F^A_{ \nu \rho}  F^{A\rho}_{\mu} $ \\
\hline
\end{tabular}
\end{center}
\end{table}
In contrast with what holds in the case of neutral DM, the covariant derivative acting on the DM now includes a coupling to the photon, which corresponds to the DM millicharge. The photon might also appear in the covariant derivative acting on the other field carrying the same millicharge as the DM  one. A third possibility for the photon to be present in the operator is through a hidden sector field strength that mixes with the hypercharge field strength. Here, the operators do not include hypercharge or $SU(2)_L$ field strengths, as they would yield a photon coupling unrelated to the DM millicharge. So the photon emission is exclusively induced by the millicharge, and will therefore be suppressed because CMB constraints on a small DM charge read \cite{Dubovsky::2001,Dolgov::2013,Dvorkin::2013}

\begin{equation}
Q^2_{DM} \leq 3.24 \times 10^{-12}  \alpha\,  (\frac{M_{DM}}{\textnormal{GeV}}) \,,
\label{mc_bound}
\end{equation}
where $\alpha$ is the electromagnetic fine structure constant.

\section{Constraints}
\label{constraints}
Once in possession of the complete list of operators describing the decay of DM into $\gamma$-ray lines, it is natural to ask whether it would be possible to discriminate among them in case of  a $\gamma$-ray line detection. To address this question, the potential indirect detection signals associated to each individual operator have been analyzed. As the operators have been studied one-by-one, potential interference effects have not been considered.

The first way to possibly discriminate among the operators is to consider the number of $\gamma$-ray lines produced through the DM decay.  Some operators might produce several lines associated to different decay channels. This type of multiple-line spectrum is  only  possible  if DM is neutral. In the case of a millicharged DM, the final state including the photon has also to include  the particle carrying the same millicharge as the DM, so there is a single possible final state for each operator. So if two lines were detected and attributed to DM decay through a single operator (the energy and intensity ratios of these lines might indicate whether this is an adequate hypothesis~\cite{neutralDDM}), that would point towards neutral DM rather than millicharged DM.

\begin{figure}[h]
\centerline{\includegraphics[width=0.50 \linewidth]{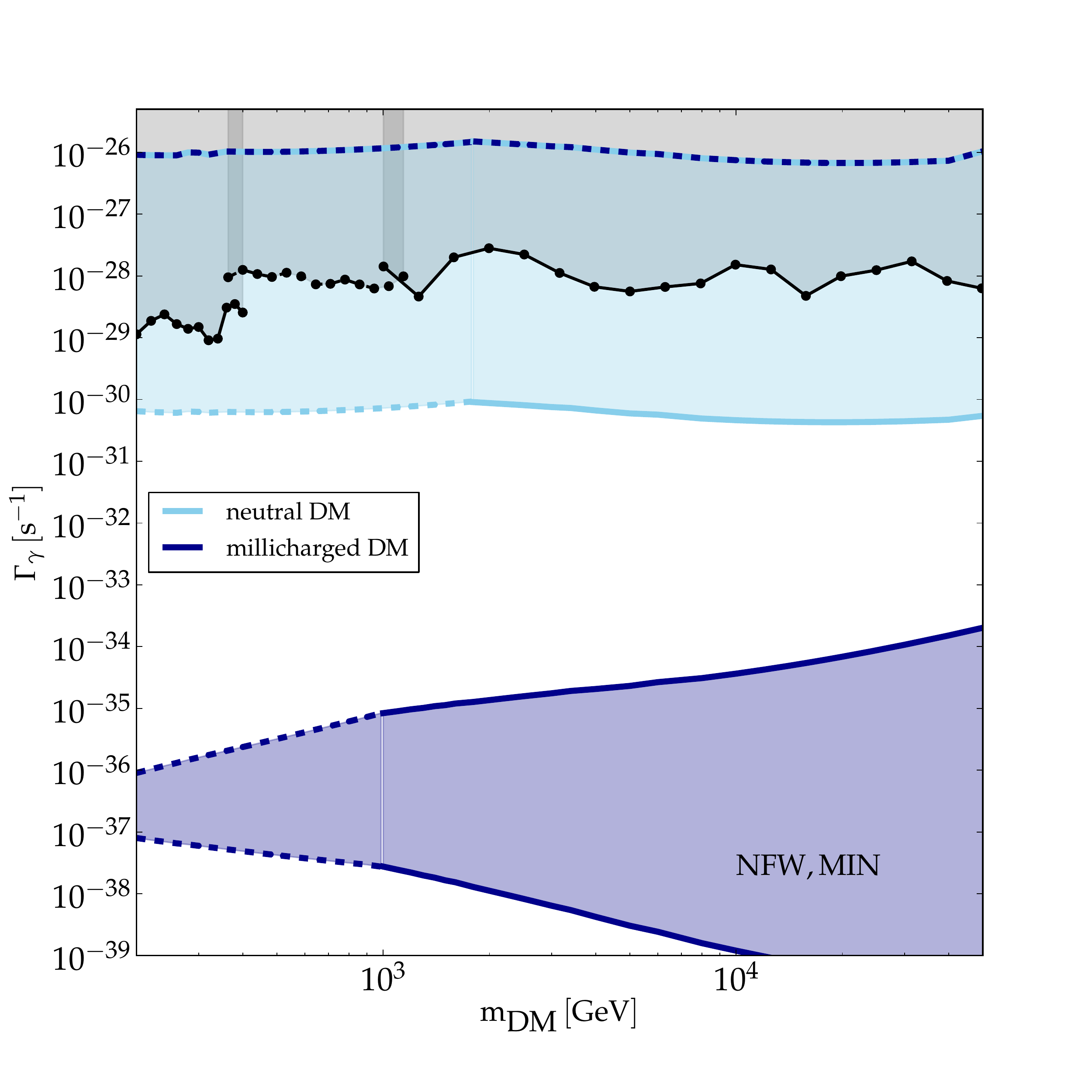}}
\caption{  Upper bounds on the decay rate into monochromatic photons from the predicted monochromatic photons over CR ratio. In light (dark) blue, the range of upper bounds obtained in the neutral (millicharged)  DM decay case. Grey areas are excluded by direct searches from Fermi-LAT and HESS experiments.  We considered a NFW profile  for the DM density and used the MIN propagation model to compute conservative $\bar{p}$  constraints \protect\cite{Cirelli::2012}.} 
\label{fig:plot}
\end{figure}

The second  mean to sort operators  out is to study the constraints deriving from continuum CR emission. When considering a DM mass above the Z mass, due to gauge invariance, the emission of a photon will always be accompanied  by the emission of at least a Z or a W boson. The SM boson will subsequently decay, producing antiprotons and continuum photons  among other particles. Depending on the quantum numbers of the DM, the coupling  to the $Z/W$ changes, and the amount of CR varies accordingly. As there are measurements of antiproton and continuum photon fluxes, it is possible to put constraints on the DM contribution \cite{Cirelli::2012} to these signals. Besides, as each operator induces a given ratio of monochromatic photons over continuum CR, this constraint can be translated into an upper bound on $\tau_{DM \rightarrow \gamma X}.$

The resuting constraints are shown in Fig. \ref{fig:plot}. In light (dark) blue, the range of upper bounds obtained in the neutral (millicharged)  DM decay case is shown. In pratice, once the quantum numbers of the DM and the companion particle are fixed, to each operator will correspond a definite photon to CR ratio, leading to an upper bound on the intensity of the $\gamma$-ray line lying in the light blue region if the DM is neutral, and in the dark blue region if it is millicharged. This regions are separated by several orders of magnitudes, except for one prediction, the dark and light blue dotted curve. This curve corresponds to the possibility of emitting a very strong $\gamma$-ray line without any excess of cosmic continuum, and it is shared by some neutral DM operators and by some millicharged DM operators, in the latter case only if the DM is a SM singlet. This is the only possibility for a millicharged DM to emit a strong line, in all the other cases a strong $\gamma$-ray line is excluded by the CR constraints. As for the neutral DM case, depending on the operator and on the quantum numbers of the particles involved, the emission of a $\gamma$-ray line with a lifetime meeting the present experimental sensitivity might be associated to an excess of CR. If both a line and a CR excess were detected, there is a potential to single out a small set of operators.

\section{Summary}
The effective theory of neutral and millicharged Dark Matter particle decaying and producing $\gamma$-ray lines has been discussed. If a line is actually detected, checking whether the spectrum exhibits one or various lines, and taking cosmic rays constraints into account might allow to distinguish the neutral from the millicharged scenario, and even to limit the compatible operators to a small subset.
\section*{Acknowledgments}

This work is supported by the FNRS-FRS and the Belgian Science Policy IAP VI-11. We thank T. Hambye for helpful  suggestions.

\end{document}